\documentclass[twocolumn,showpacs,amsmath,amssymb]{revtex4-1}
\usepackage{graphicx}

\renewcommand{\figurename}{Fig.}

\begin{document}

\title{Information Geometry for Husimi-Temperley Model}

\author{Yoichiro Hashizume}
 \email{hashizume@rs.tus.ac.jp}
\affiliation{Department of Applied Physics, Tokyo University of Science, Tokyo 125-8585, Japan}
 
\author{Hiroaki Matsueda}%
 \email{matsueda@sendai-nct.ac.jp}
\affiliation{Sendai National College of Technology, Sendai 989-3128, Japan}%
\date{\today}
\begin{abstract}
We examine phase transition of the Husimi-Temperley model in terms of information geometry. For this purpose, we introduce the Fisher metric defined by the density matrix of the model. We find that the metric becomes hyperbolic at the critical point with respect to the energy scale. Then, the metric is invariant under the scale transformation. We also find that the equation of states is naturally derived from a necessary condition for the entropy operator that is a building block of the metric. Based on these findings, we conclude that the geometric quantities clearly detect the phase transition of the model.
\end{abstract}
\maketitle

\section{Introduction}

Geometrical approaches to current physics are recognized to be quite powerful tools. Although differential-geometrical viewpoints have long stories in general relativity, recent two majorities are holographic and topological viewpoints. Geometry picks up invariant quantities, and thus the quantities characterize phases of matters in terms of universality. This indicates that the theory for critical phenomena can be translated into a simple geometric description.

In condensed matter physics and statistical physics, it is a challenging theme to apply holographic principles to sophisticated analysis of various models that are not easily treated within standard approaches. Starting from a complicated microscopic model, we consider how we transform it into a geometric or macroscopic description, in which tracing over microscopic degrees of freedom makes it possible to understand the model in much simpler terminology. One of key holographic principles is the so-called anti-de Sitter space / conformal field theory (AdS/CFT) correspondence~\cite{AdS/CFT}. This correspondence was originally developed in superstring theory, but a current topic is to examine how to apply this correspondence to much broader physical problems. In practice, the multiscale entanglement renormalization ansatz (MERA) and the Ryu-Takayamagi formula play key roles on these examinations~\cite{MERA,Swingle,HMss,f-TMERA,RT}.

In this paper, we establish an information geometrical treatment of phase transitions for statistical models, and give a simple example based on the Husimi-Temperley model\cite{Husimi, Temp}. This is an alternative way to transform a statistical model into a geometric description. We particularly focus on the information space into which the overall data of the partition function of the original model are embedded by an appropriate way. Then, the model parameters such as temperature and magnetic field correspond to the space coordinates in the information space, and the magnitude of the entropy can be viewed as the field strength in the information space. Furthermore, the measure in the space is defined by the Fisher metric (or the relative entropy) for the Boltzmann distribution~\cite{Fisher}. In general the space is not flat, and we would like to calculate the metric on the information space. According to one-to-one correspondence between an invariant quantity such as curva
 ture in the geometrical description and criticality in the original model, we can expect that the curvature would pick up fundamental information of phase transition. This is because the second-order phase transition is characterized by the conformal symmetry and this symmetry matches well with the hyperbolic geometry with negative curvature. Actually, the hyperbolic metric is invariant under the conformal transformation. Radically saying, the hyperbolic metric is a good signal for detecting the second-order phase transition. The main objective of this paper is to confirm this conjecture.

We will find that the Fisher metric for the Husimi-Temperley model becomes hyperbolic at the critical point with respect to the energy scale. Although this is a mean-field model, we expect that the hyperbolic nature characterizes scale invariance at criticality. We will also find that the equation of states is derived from a necessary condition for the metric. Based on these observations, we will conclude that the geometric quantities have enough information of the phase transition of the model.

This paper is organized as follows: In the following section, we introduce a representation of Fisher metric using the density matrix which is well-defined even in quantum systems.
In section III, we obtain the Fisher metric of Husimi-Temperley model\cite{Husimi,Temp}, which shows a hyperbolic structure with respect to energy scale at the critical point.
Summary is included in section IV.

\section{Fisher Metric}

Let us first define the Fisher metric for general statistical models. For later convenience, we generalize the standard definition of the Fisher metric so that the method can be easily applicable to any quantum systems.

The density matrix $\rho$ of the equilibrium states for the Hamiltonian $\mathcal{H}$ is defined as
\begin{equation}
\rho=\frac{1}{Z(\beta)}e^{-\beta\mathcal{H}}=e^{-\beta(\mathcal{H}-F)},\label{eq2-1}
\end{equation}
where the parameters $\beta, Z(\beta)$ and $F$ denote the inverse temperature $\beta=1/k_{\text{B}}T$, the partition function $Z(\beta)={\text{Tr}}\exp [-\beta \mathcal{H}]$ and the free energy $F=-k_{\text{B}}T\log Z(\beta)$, respectively.
Using the density matrix $\rho$, we define the matrix $\gamma (\rho)$ as
\begin{equation}
\gamma (\rho)=-\log \rho=\beta\mathcal{H}+\log Z(\beta),\label{eq2-2}
\end{equation}
which is called as entropy operator~\cite{Zubentropy,MSneentropy2,MSneentropy1}. The expectation value of $\gamma(\rho)$ is equal to the thermal entropy, $S=k_{\text{B}}\left<\gamma(\rho)\right>$. According to the identity
\begin{equation}
\frac{d}{d\xi}e^{A(\xi)}=\int_{0}^{1}d\lambda e^{(1-\lambda)A(\xi)}\left( \frac{d A(\xi)}{d\xi}\right)e^{\lambda A(\xi)}\label{eq2-3}
\end{equation}
for a matrix $A(\xi)$ which depends on the c-number parameter $\xi$, we can show the mean values of the differentials of $\gamma (\rho)$ vanish, namely $\langle \partial_{\mu} \gamma (\rho) \rangle=0$, as follows:
\begin{align}
\frac{d}{dx^{\mu}}\rho &= \int_{0}^{1}d\lambda \rho^{(1-\lambda)}\left( \frac{d (\log \rho)}{dx^{\mu}}\right) \rho^{\lambda} \notag
\\
\Rightarrow \frac{d}{dx^{\mu}}\text{Tr}\rho &= \int_{0}^{1}d\lambda {\text{Tr}} \rho \left( \frac{d (\log \rho)}{dx^{\mu}}\right) \notag
\\
\Leftrightarrow  0 &= \langle \partial_{\mu} \gamma (\rho) \rangle. \label{eq2-4}
\end{align}
Here we use the normalization condition ${\text{Tr}}\rho =1$ and we take $A(\xi)$ as $\log \rho$.

Furthermore, one more differentiation of Eq.(\ref{eq2-4}) by $x^{\nu}$ leads to
\begin{align}
0 &= \partial_{\nu} \left( {\text{Tr}}\rho \partial_{\mu} \gamma(\rho) \right) \notag
\\
&={\text{Tr}} (\partial_{\nu} \rho)\partial_{\mu}\gamma (\rho) + {\text{Tr}} \rho \partial_{\nu}\partial_{\mu}\gamma(\rho).\label{eq2-5}
\end{align}
Then we can obtain the relation
\begin{align}
\langle \partial_{\nu}\partial_{\mu} \gamma(\rho) \rangle &=-{\text{Tr}}(\partial_{\nu}\rho)(\partial_{\mu}\gamma (\rho)) \notag
\\
&= -{\text{Tr}}\int^{1}_{0}d\lambda \rho^{(1-\lambda)}\left(\partial_{\nu}\log \rho\right)\rho^{\lambda}\left(\partial_{\mu}\gamma (\rho)\right)\notag
\\
&= {\text{Tr}}\int^{1}_{0}d\lambda \rho e^{\lambda \beta\mathcal{H}}\left(\partial_{\nu}\gamma (\rho) \right)e^{-\lambda \beta \mathcal{H}}\left(\partial_{\mu}\gamma (\rho)\right) \notag
\\
&= \frac{1}{\beta} \int^{\beta}_{0} d\lambda {\text{Tr}}\rho e^{\lambda \mathcal{H}}\left(\partial_{\nu}\gamma (\rho) \right)e^{-\lambda \mathcal{H}}\left(\partial_{\mu}\gamma (\rho)\right) \notag
\\
&= \langle \partial_{\nu} \gamma (\rho);\partial_{\mu} \gamma (\rho) \rangle, \label{eq2-6}
\end{align}
using Eqs.(\ref{eq2-2}) and (\ref{eq2-3}).
The final representation $\langle \partial_{\nu} \gamma (\rho);\partial_{\mu} \gamma (\rho) \rangle$ in Eq.(\ref{eq2-6}) is so called canonical correlation between $\partial_{\nu} \gamma (\rho)$ and $\partial_{\mu} \gamma (\rho)$~\cite{Kubo}.
This is equal to the classical correlation $\langle \partial_{\nu} \gamma (\rho) \partial_{\mu} \gamma (\rho) \rangle$, when $e^{-\lambda{\cal H}}$ and $\partial_{\nu}\gamma(\rho)$ commute with each other. This classical correlation is nothing but the standard Fisher metric~\cite{Fisher}.
Then we can generally define the metric tensor $g_{\mu \nu}$ for statistical systems as 
\begin{equation}
g_{\mu \nu}\equiv \langle \partial_{\mu} \gamma (\rho);\partial_{\nu} \gamma (\rho) \rangle = \langle \partial_{\mu}\partial_{\nu} \gamma(\rho) \rangle \label{eq2-7}
\end{equation} 
using the density matrix $\rho$.

\section{Geometrical analysis of Husimi-Temperley model}

In the previous section, we have discussed general representation of the metric tensor $g_{\mu\nu}$.
In this section, we apply it to the examination of statistical properties of the Husimi-Temperley model at the critical point.

\subsection{Husimi-Temperley model}

The Husimi-Temperley model is defined by the following Hamiltonian\cite{Husimi,Temp}:
\begin{equation}
\mathcal{H}=-\frac{J}{2N}\sum_{i \not= j}S_iS_j. \label{eq3-1}
\end{equation}
All the spins are mutually interacting with the same exchange coupling $J$. Thus the mean-field solution becomes exact. This example might be too simple for our present purpose, but at the same time this simple example provides us very clear overall structure of our setup. Therefore, more comlicated and realistic models would be future targets. The density matrix $\rho=\exp[-\beta(\mathcal{H}-F)]$ yields the metric tensor $g_{\mu\nu}$ as
\begin{align}
g_{\mu\nu}
&=\langle \partial_{\mu}\partial_{\nu}(-\log \rho) \rangle \notag \\
&=\langle \partial_{\mu}\partial_{\nu}(\beta\mathcal{H}) \rangle+\partial_{\mu}\partial_{\nu}\log Z(K,m), \label{eq3-2}
\end{align}
where the parameters $K$ and $m$ denote the normalized interaction $K=\beta J$ and the single spin moment $m=\langle S_i\rangle$, respectively.
Here the parameters $x^{\mu}$ ($\mu=0,1$) are defined as $x^0=K$ and $x^1=m$.
As shown in the following discussion, the magnetization $m$ depends on the temperature $T$ as well as the parameter $K$.
However, at this stage, we treat them as independent parameters, and their relation will be introduced as a constraint.
For these assumptions, we know that the first term in Eq.~(\ref{eq3-2}) vanishes, since the Hamiltonian (\ref{eq3-1}) does not include $m$ explicitly and is only a linear function of $K$. As a result, the metric tensor $g_{\mu\nu}$ is obtained as
\begin{equation}
g_{\mu\nu}= \partial_{\mu}\partial_{\nu} \log Z(K,m).\label{eq3-3}
\end{equation}

As is well known, the function $\log Z(K,m)$ for the Husimi-Temperley model is obtained as
\begin{equation}
\log Z(K,m)=-\frac{NKm^2}{2}+N\log 2\cosh Km.\label{eq3-4}
\end{equation}
Thus, using the relations
\begin{equation}
\partial_{K}\log Z(K,m)=-\frac{Nm^2}{2}+Nm\tanh Km,\label{eq3-5}
\end{equation}
and
\begin{equation}
\partial_{m}\log Z(K,m)=-NKm+NK\tanh Km,\label{eq3-6}
\end{equation}
we obtain the metric tensor $g_{\mu\nu}$ as
\begin{align}
g_{KK}&= \partial_{K}\partial_{K}\log Z(K,m)=\frac{Nm^2}{\cosh^2 Km},\label{eq3-7}
\end{align}
\begin{align}
g_{Km}=g_{mK}&= \partial_{m}\partial_{K}\log Z(K,m)\notag
\\
&=-Nm+\frac{NmK}{\cosh^2 Km}+N\tanh Km,\label{eq3-8}
\end{align}
and
\begin{align}
g_{mm}&= \partial_{m}\partial_{m}\log Z(K,m)=-NK+\frac{NK^2}{\cosh^2 Km}.\label{eq3-9}
\end{align}
Then the line element $ds^2$ is represented as
\begin{align}
ds^2=&g_{\mu\nu}dx^{\mu}dx^{\nu}\notag
\\
=&\frac{Nm^2}{\cosh^2 Km}dK^2\notag
\\
&+2N\left( \frac{mK}{\cosh^2 Km}+\tanh Km -m\right) dKdm\notag
\\
&+N\left( \frac{K^2}{\cosh^2 Km}-K\right)dm^2.\label{eq3-10}
\end{align}
As shown in the previous section, the expectation values $\langle \partial_{m}\gamma(\rho) \rangle$ and $\langle \partial_{K}\gamma(\rho) \rangle$ vanish. For the Husimi-Tempeley model, they are respectively equal to
\begin{eqnarray}
\partial_{m}\log Z(K,m)=0,
\end{eqnarray}
and
\begin{equation}
-\frac{1}{2N}\sum_{i\ne j}\langle S_{i}S_{j}\rangle + \partial_{K}\log Z(K,m) =0.
\end{equation}
Each of them leads to the following equation:
\begin{equation}
m=\tanh Km.\label{eq3-11}
\end{equation}
This is nothing but the equation of states to determine the $T$ dependence of $m$. Furthermore, the relation between $dm$ and $dK$ is derived as
\begin{align}
&dm=\partial_{K} (\tanh Km)dK +\partial_{m} (\tanh Km)dm\notag
\\
&\Leftrightarrow \left(1-\frac{K}{\cosh^2 Km}\right) dm =\frac{m}{\cosh^2 Km}dK\notag
\\
&\Leftrightarrow dm=\frac{m}{\cosh^2 Km -K}dK.\label{eq3-12}
\end{align}
Inserting Eqs.(\ref{eq3-11}) and (\ref{eq3-12}) into Eq.(\ref{eq3-10}), the line element $ds^2$ is represented as
\begin{align}
ds^2=&\frac{Nm^2}{\cosh^2 Km}dK^2+2\frac{NmK}{\cosh^2 Km}dKdm\notag
\\
&-N\left( K- \frac{K^2}{\cosh^2 Km}\right) dm^2\notag
\\
=&\frac{Nm^2}{\cosh^2 Km}dK^2+NK\left( 1- \frac{K}{\cosh^2 Km} \right) dm^2\notag
\\
=&\frac{Nm^2}{\cosh^2 Km-K}dK^2.\label{eq3-13}
\end{align}
Here the denominator in Eq.(\ref{eq3-13}) is approximated as
\begin{equation}
\cosh^2 Km -K\sim (1-K)+ (Km)^2\sim (Km)^2, \label{eq3-14}
\end{equation}
near the critical point, $K\sim K_{\text{c}}=1$ and $m\sim 0$.
Then, the line element $ds^2$ shown in Eq.(\ref{eq3-13}) yields the asymptotic form as
\begin{equation}
ds^2\sim N\frac{dK^2}{K^2}.\label{eq3-15}
\end{equation}
Mathematically, any one-dimensional metric is transformed into Euclidean by general coordinate transformation. However, if we focus on the fact that the parameter $K$ represents the energy scale of the model, the hyperbolic nature of Eq.~(\ref{eq3-15}) is still suggestive for understanding how this parameter $K$ controls the nature of the phase transition in the information space. This metric is clearly scale invariant with respect to the energy scale $K$. Of course, the universality class of the present model is in mean-field type, and it is hard to represent fractal spin configuration like the Ising model. However, in group-theoretical viewpoints like those in the AdS/CFT correspondence, the scale invariance in the information space seems to match with the criticality of the second-order phase transition in the original model. In order to confirm stability of the hyperbolic metric, it would be necessary to examine time dependence of the Husimi-Temperley model and to examine
  whether we can obtain $ds^{2}\sim(d\tau^{2}+dK^{2})/K^{2}$ for the normalized time $\tau$.


\subsection{Husimi-Temperley model with external field}

Let us consider how the external field $H$ contribute to the line element $ds^2$.
The external (or magnetic) field $H$ affects to spins on Husimi-Temperley model as shown in the Hamiltonian
\begin{equation}
\mathcal{H}=-\frac{J}{2N}\sum_{i\not=j}S_iS_j-\mu_{\text{B}}H\sum_{j}S_j.\label{eq3-16}
\end{equation}
In the present case, the function $\log Z(K,h,m)$ is obtained as
\begin{equation}
\log Z(K,h,m)=-\frac{NKm^2}{2}+N\log 2\cosh (Km+h),\label{eq3-17}
\end{equation}
where the parameter $h$ denotes the normalized external field, namely $h=\beta\mu_{\text{B}}H$.
Thus, $\gamma(\rho)$ is given by
\begin{equation}
\gamma (\rho)=-\frac{K}{2N}\sum_{i\not=j}S_iS_j-h\sum_{j}S_j+\log Z(K,h,m),\label{eq3-18}
\end{equation}
and this leads to the metric tensor $g_{\mu\nu}$ using the parameters $K,h$ and $m$ as
\begin{align}
g_{KK}&=\partial_{K}\partial_{K}\gamma(\rho)= \frac{Nm^2}{\cosh^2(Km+h)},\label{eq3-19}
\\
g_{Kh}&=g_{hK}=\partial_{K}\partial_{h}\gamma(\rho)=\frac{Nm}{\cosh^2(Km+h)},\label{eq3-20}
\\
g_{Km}&=g_{mK}=\partial_{K}\partial_{m}\gamma(\rho) \notag
\\
&=-Nm+N\tanh(Km+h)+\frac{NKm}{\cosh^2(Km+h)},\label{eq3-21}
\\
g_{hh}&=\partial_{h}\partial_{h}\gamma(\rho)=\frac{N}{\cosh^2(Km+h)},\label{eq3-22}
\\
g_{hm}&=g_{mh}=\partial_{h}\partial_{m}\gamma(\rho)=\frac{NK}{\cosh^2(Km+h)},\label{eq3-23}
\end{align}
and
\begin{align}
g_{mm}=\partial_{m}\partial_{m}\gamma(\rho)=-NK+\frac{NK^2}{\cosh^2(Km+h)}.\label{eq3-24}
\end{align}
Similarly to the previous subsection, the differential of $\gamma(\rho)$ satisfying Eq.(\ref{eq2-4}) derives the following equation of states;
\begin{equation}
\langle \partial_{\mu}\gamma (\rho) \rangle=0 \Leftrightarrow m=\tanh(Km+h). \label{eq3-25}
\end{equation}
The relation between $dK, dh$ and $dm$ is obtained by Eq.(\ref{eq3-25}) as
\begin{align}
dm=&\partial_{K} (\tanh (Km+h))dK \notag
\\
&+\partial_{h} (\tanh (Km+h))dh+\partial_{m} (\tanh (Km+h))dm\notag
\\
&\Leftrightarrow dm=\frac{mdK+dh}{\cosh^2(Km+h)-K}.\label{eq3-26}
\end{align}
Then, the line element $ds^2$ is obtained as
\begin{equation}
ds^2=g_{\mu\nu}dx^{\mu}dx^{\nu}=\frac{N(mdK+dh)^2}{\cosh^2(Km+h) -K}.\label{eq3-27}
\end{equation}
Now we consider the behavior of $ds^2$ near the critical point described by the conditions $K\sim K_{\text{c}}=1, h\sim 0$ and $m\sim 0$.
The equation (\ref{eq3-27}) is approximated into
\begin{equation}
ds^2\sim N\left(\frac{mdK+dh}{mK+h}\right)^2. \label{eq3-28}
\end{equation}
The result means that the external field facilitates the magnetic order at higher energy scale. This is also consistent with the standard analysis.

\section{Summary}

Summarizing, we have examined phase transition of the Husimi-Temperley model in terms of Fisher metric. Our message is that the metric clearly represents the nature of the phase transition. In particular, the metric is invariant under the energy-scale transformation near the critical point. The invariance is related to the criticality of the second-order phase transition. Furthermore, we have found that the necessary condition for the metric leads to the equation of states. Future study is to go beyond mean-field level. It is also important to examine time dependence of the Husimi-Temperley model.

Finally, we compare the present approach with the AdS/CFT correspondence. In the present case, both of classical and quantum systems can be considered, and this concept would be much broader than that of the AdS/CFT correspondence.
The Husimi-Temperley model is mean-field one, and thus there is no typical length scale and spatial axis. Then, the physics is determined by only the local moment. In this sense, we may say that the model is zero-dimensional one (Note that this situation is called as `infinite dimensional' in statistical mechanics, since a local spin interacts with the infinite number of spins). In the local model, non-local quantum correlation is in principle nothing. Therefore, once we recognize quantum correlations need not to be strictly distinguished with classical one, the emergence of the hyperbolic nature might be regarded as the lowest-dimensional toy model of the AdS/CFT correspondence.

\acknowledgements
The work of Y. H. was financially supported by JSPS Grant-in-Aid for Young Scientists (B) Grant Number 26800205.
Y. H. wishes to thank Prof. S. Okamura (TUS) for kind supports, particularly discussing some possible applications.


\begin{thebibliography}{00}

\bibitem{AdS/CFT}
J. Maldacena, Adv. Theor. Math. Phys. {\bf 2}, 231 (1998).

\bibitem{MERA}
G. Vidal, Phys. Rev. Lett. {\bf 99}, 220405 (2007).
\bibitem{Swingle}
B. Swingle, Phys. Rev. D {\bf 86}, 065007 (2012).
\bibitem{HMss}
H. Matsueda, Phys. Rev. E {\bf 85}, 031101 (2012). 
\bibitem{f-TMERA}
H. Matsueda, M. Ishihara, Y. Hashizume, Phys. Rev. D {\bf 87}, 066002 (2013).    
\bibitem{RT}
S. Ryu and T. Takayanagi, Phys. Rev. Lett. {\bf 96}, 181602 (2006).

\bibitem{Husimi}
K. Husimi, Proc. Int. Conf. Theor. Phys. {\bf 531} (1953).
\bibitem{Temp}
H. N. Temperley, Proc. Phys. Soc. {\bf 67}, 233 (1954).

\bibitem{Fisher}
R. A. Fisher, Proc. Cambridge Philos. Soc. {\bf 22}, 700 (1925).

\bibitem{Zubentropy}
D. N. Zubarev, {\it ``Nonequilibrium Statistical Mechanics''}, Nauka, 1971, Moscow.
\bibitem{MSneentropy2}
M. Suzuki, Int. J. Mod. Phys. B {\bf 10} (1996) 1637.
\bibitem{MSneentropy1}
M. Suzuki, Physica A {\bf 391} (2012) 1074.

\bibitem{Kubo}
R. Kubo, J. Phys. Soc. Jpn. {\bf 12} (1957) 570.

\end{thebibliography}
\end{document}